\documentclass[twoside,12pt]{article}
\usepackage{amssymb}
\usepackage{graphicx}
\usepackage{amsmath}
\usepackage{amsfonts}

\newtheorem{theorem}{Theorem}

\newtheorem{proposition}{Proposition}

\begin{document}
\noindent
\textbf{\Large Characterizations Of Umbilically Synchronized Space-times}\\ 
\noindent
\vskip.1in
\noindent
Ramesh Sharma\\
Department of Mathematics, University of New Haven, CT 06516, USA\\

\noindent
E-mail: rsharma@newhaven.edu\\

\noindent
\textbf{Abstract}\\
We study umbilically synchronized space-times $M$ that have vanishing electric part of the Weyl tensor and show that  (i) If $dim(M) = 4$, then $M$ is conformally flat, (ii) If $dim(M) > 4$, $M$ is conformally flat if and only if the spatial slices are conformally flat. Next, we characterize umbilically synchronized spacetimes that are vacuum and have zero electric components of its Weyl tensor. Finally, we characterize conformal vector fields of an umbilically synchronized space-time.\\

\noindent
\textit{AMS Subj. Class.}: 53 C50,53 B21, 83 C20.\\


\section{Introduction} The standard Friedmann-Lemaitre-Robertson-Walker ($FLRW$) cosmological model is described by the space-time $M$ with the line-element

\begin{equation}\label{1}
ds^2 = g_{\alpha \beta}dx^{\alpha}dx^{\beta}=-dt^2 +a^{2}(t)\gamma_{ij}dx^{i}dx^{j}
\end{equation}
where $\alpha,\beta$ denote the space-time indices running over 0,1,2,3, the spatial indices $i,j$ run over 1,2,3, time coordinate $t = x^0$, the warping function $a(t)$ is the scale function, and $\gamma_{ij}$ is the fixed spatial metric of constant curvature. We know that (i) $M$ is conformally flat, (ii) the spatial slices $\Sigma$ ($t$ = constant) have constant intrinsic curvature and are homothetic to one another, and (iii) $\Sigma$ are umbilical in $M$ and have constant mean curvature. A general ($n+1$)-dimensional space-time is described in the $ADM$ (Misner, Thorne and Wheeler \cite{M-T-W}) formalism by the metric $g_{\alpha \beta}$ such that the line-element is
\begin{equation}\label{2}
ds^2 = g_{\alpha \beta}dx^{\alpha}dx^{\beta}= -N^2 dt^2+g_{ij}(dx^i + S^i dt)(dx^j +S^j dt)
\end{equation}
where the greek indices $\alpha, \beta$ run over 0,1,...,n and latin indices $i,j$ over 1,...,n; $N$ is the Lapse function that depends on $t$ and $x^i$, and represents the clock rates for an observer relative to a reference system of clocks, and $S^i$ is a vector field on the $n$-dimensional slice $\Sigma$ ($t$ = constant) which represents two observers in relative motion with velocity $S^i$. In this paper, we assume that the shift vector $S^i$ is zero, i.e. the evolution vector field $\frac{\partial}{\partial t}$ is orthogonal to the spatial slices $\Sigma$ and also that $\Sigma$ are totally umbilical in $M$ (which are true for the $FLRW$ spacetime). However, the mean curvature need not be constant on any slice. Space-times foliated by such $\Sigma$'s are called umbilically synchronized space-times (see Ferrando, Morales and Portilla \cite{F-M-P}) and are shear-free and vorticity-free with respect to an observer whose congruence is given by the unit vector $\textbf{n} = \frac{1}{N}\frac{\partial}{\partial t}$ normal to $\Sigma$. The acceleration vector field $\textbf{A} = \bar {\nabla}_{\bf{n}}\bf{n}$ need not vanish. Treciokas and Ellis \cite{T-E} have shown that the shear-free and vorticity-free timelike congruences constitute a large class among the observers measuring an isotropic distribution function obeying the Boltzmann equation. Conformally flat umbilical synchronizations exist in any space-time admitting natural symmetric frames (Coll and Morales \cite{C-M}). This occurs for any spherically symmetric space-time and also for the Stephani universes that include the $FLRW$ space-time.\\

\noindent
In this paper we study umbilically synchronized space-times ($M,g$) and derive the components of its Weyl conformal tensor in terms of the geometric quantities of spatial slices $\Sigma$ ($t =$ constant). Then we obtain a condition for ($M,g$) to be conformally flat in terms of the electric components of the Weyl tensor and conformal flatness of the spatial slices. Next we characterize umbilically synchronized spacetimes that are vacuum and have zero electric components of its Weyl tensor. Finally, we provide some characterizations of conformal (including inheriting) vector fields of an umbilically synchronized space-time.

\section{Basic Equations} We denote the Levi-Cita connection, the Riemann curvature tensor, Ricci tensor, scalar curvature and the Weyl tensor of the metric $g_{ij}$ by $\nabla$, $R_{ijkl}$, $R_{ij}$, $R$ and $C_{ijkl}$. Corresponding quantities of the space-time metric $g_{\alpha \beta}$ are denoted by bars over the corresponding symbols with greek indices $\alpha, \beta, \gamma, \delta$ in lieu of the latin indices $i,j,k,l$. As indicated earlier, the unit vector field $\bf{n}$ $= \frac{1}{N}\frac{\partial}{\partial t}$ is normal to $\Sigma$, and the acceleration vector field $\textbf{A} = \bar {\nabla}_{\bf{n}}\bf{n}$ is tangential to the space-like slices $\Sigma$, and can be shown by a direct computation, to be equal to the spatial gradient of $ln N$. Denoting the co-ordinate basis of the tangent space of $\Sigma$ by $\partial_i$ we have the second fundamental form $K_{ij}$ of $\Sigma$ defined by $g(\bar{\nabla}_{\partial_i}\bf{n},\partial_j)$, where $g$ is the space-time metric. The umbilicity of $\Sigma$ in $M$ means that
\begin{equation}\label{3}
K_{ij}=\tau g_{ij}
\end{equation}
where $\tau$ stands for the mean curvature of $\Sigma$. It can be verified by an easy computation (see Fischer and Marsden  \cite{F-M}) that 
\begin{equation}\label{4}
K_{ij}=\frac{1}{2N}\partial_t g_{ij}
\end{equation}
Equations (\ref{3}) and (\ref{4}) imply the linear differential equation $\partial_t g_{ij} = 2N\tau g_{ij}$ whose solution $g_{ij}$ splits off a time-independent metric $\gamma_{ij}$ such that
\begin{equation}\label{5}
g_{ij}=a^2 (t,x^k)\gamma_{ij}
\end{equation}
for a positive function $a$ that depends on $t$ and $x^k$. Thus the line-element of the umbilically synchronized space-time assumes the form:
\begin{equation*}
-N^2 dt^2 +a^2(t,x^k)\gamma_{ij}dx^i dx^j.
\end{equation*}
In particular, for $N=1$ and $a$ a function of only $t$ and $\gamma$ any fixed time-independent Riemannian metric, we get generalized Robertson-Walker space-time (Alias, Romero and Sanchez \cite{A-R-S}) which become $FLRW$ space-time when $\gamma$ is of constant curvature. Comparing equation (\ref{3}) with (\ref{4}) and using (\ref{5}) yields the relation
\begin{equation}\label{6}
\tau=\frac{\dot a}{aN}
\end{equation}
where the over-dot denotes partial differentiation with respect to $t$. The classical Gauss and Codazzi equations for $\Sigma$ are:
\begin{equation}\label{7}
\bar{R}_{ijkl}=R_{ijkl}+K_{il}K_{jk}-K_{ik}K_{jl}
\end{equation}
\begin{equation}\label{8}
\bar{R}_{ijk0}=\nabla_j K_{ki}-\nabla_i K_{kj}
\end{equation}
We also have the following mixed components as given in \cite{F-M}:
\begin{equation}\label{9}
\bar{R}_{0i0j}=N^2 (\frac{1}{N}\partial_t K_{ij}-K^k_i K_{kj}-\frac{1}{N}\nabla_i \nabla_j N)
\end{equation} 
Next, using the umbilicity condition (\ref{3}) and the above curvature components in the definition $\bar{R}_{\alpha \beta} = g^{\gamma \delta}\bar{R}_{\gamma \alpha \beta \delta}$ we obtain 

\begin{equation}\label{10}
\bar{R}_{ij}=R_{ij}+(n\tau^2 +\frac{\dot \tau}{N})g_{ij}-\frac{1}{N}\nabla_i \nabla_j N
\end{equation}
\begin{equation}\label{11}
\bar{R}_{i0}=(1-n)\nabla_i \tau
\end{equation}
\begin{equation}\label{12}
\bar{R}_{00}=N(\Delta N -Nn\tau^2 -n\dot \tau)
\end{equation}
\begin{equation}\label{13}
\bar{R}=R+2\frac{n}{N}\dot \tau +n\tau^2+n^2\tau^2-\frac{2}{N}\Delta N
\end{equation}
At this point we recall that the Weyl conformal tensor $\bar{C}$ of the (n+1)-dimensional space-time is given by the components
\begin{eqnarray*}
\bar{C}_{\alpha \beta \gamma \delta}&=&\bar{R}_{\alpha \beta \gamma \delta}-\frac{1}{n-1}(\bar{R}_{\beta \gamma}g_{\alpha \delta}\nonumber-\bar{R}_{\alpha \gamma}g_{\beta \delta}+g_{\beta \gamma}\bar{R}_{\alpha \delta}\nonumber\\
&-&g_{\alpha \gamma}\bar{R}_{\beta \delta})+\frac{\bar{R}}{n(n-1)}(g_{\beta \gamma}g_{\alpha \delta}-g_{\alpha \gamma}g_{\beta \delta})
\end{eqnarray*}
Using this definition along with equations (\ref{7})-(\ref{9}) and (\ref{10})-(\ref{13}), and after a lengthy computations and arrangements, we obtain

\begin{eqnarray}\label{14}
\bar{C}_{ijkl}&=&C_{ijkl}+\frac{1}{(n-1)(n-2)}[g_{il}(R_{jk}-\frac{R}{n}g_{jk})-g_{jl}(R_{ik}-\frac{R}{n}g_{ik})\nonumber\\
&+&g_{jk}(R_{il}-\frac{R}{n}g_{il})-g_{ik}(R_{jl}-\frac{R}{n}g_{jl})]+\frac{1}{N(n-1)}[g_{il}(\nabla_j \nabla_k N\nonumber\\
&-&\frac{\Delta N}{n}g_{jk})-g_{jl}(\nabla_k \nabla_i N-\frac{\Delta N}{n}g_{ki})+g_{jk}(\nabla_i \nabla_l N\nonumber\\
&-&\frac{\Delta N}{n}g_{il})-g_{ik}(\nabla_j \nabla_l N-\frac{\Delta N}{n}g_{jl}]
\end{eqnarray}

\begin{equation}\label{15}
\bar{C}_{ijk0}=0
\end{equation}

\begin{equation}\label{16}
\bar{C}_{0i0j}=-\frac{N^2}{n-1}[R_{ij}-\frac{R}{n}g_{ij}+\frac{n-2}{N}(\nabla_i \nabla_j N-\frac{\Delta N}{n}g_{ij})]
\end{equation} 

\section{A Conformal Flatness Criterion}
Let us recall (Stephani et al. \cite{Stephani}) that the electric part of the Weyl tensor of the space-time with respect to $\bf{n}$ is $E_{ij} = \bar{C}(\partial_i,\bf{n},\partial_j,\bf{n})=$ $\frac{1}{N^2}\bar{C}_{i0j0}$. So, if $E_{ij} = 0$, then equation (\ref{16}) provides
\begin{equation}\label{17}
R_{ij}-\frac{R}{n}g_{ij}+\frac{n-2}{N}(\nabla_i \nabla_j N-\frac{\Delta N}{n}g_{ij})=0
\end{equation}
If $M$ is 4-dimensional, then $\Sigma$ is 3-dimensional and hence $C_{ijkl} = 0$. Using this in (\ref{14}) shows that $\bar{C}_{ijkl} = 0$ and hence $M$ is conformally flat. For $\textnormal{dimension } M > 4$, equations (\ref{14}) and (\ref{17}) imply $\bar{C}_{ijkl}=C_{ijkl}$, and so if $C_{ijkl}=0$, i.e. $\Sigma$'s are conformally flat, then $\bar{C}_{ijkl}=0$, and hence $M$ is conformally flat. Converse is evident. We state these outcomes as the following result.
\begin{theorem}Let the electric part of the Weyl tensor of an umbilically synchronized space-time $M$ of dimension $\ge 4$, be zero. If $dim(M) = 4$, then $M$ is conformally flat. For $dim(M) > 4$, $M$ is conformally flat if and only if $\Sigma$'s are conformally flat.
\end{theorem}

\noindent
\textbf{Remark.} This result is a generalization of the classical result \cite{Sharma-Duggal} ``A generalized Robertson-Walker space-time $M$ with metric: $-dt^2 + a^{2}(t)\gamma_{ij}dx^i dx^j$ (where $\gamma$ is a time-independent Riemannian metric) is conformally flat if and only if $\gamma_{ij}$ has constant curvature (and hence conformally flat). More generally, the electric part of the Weyl tensor of $M$ is zero if and only if $\gamma_{ij}$ is Einstein."

\section{Vacuum Case}
For umbilically synchronized space-times that are vacuum, i.e. $\bar{R}_{\alpha \beta}=0$, equations (\ref{10}), (\ref{11}), (\ref{12}) and (\ref{13}) provide
\begin{equation}\label{18}
R_{ij}-\frac{R}{n}g_{ij}=\frac{1}{N}(\nabla_i \nabla_j N-\frac{\Delta N}{n}g_{ij})
\end{equation}
\begin{equation}\label{19}
(1-n)\nabla_i \tau = 0
\end{equation}
\begin{equation}\label{20}
\Delta N = n(\dot \tau +N\tau^2)
\end{equation}
\begin{equation}\label{21}
R=-2\frac{n}{N}\dot \tau -n\tau^2-n^2\tau^2+\frac{2}{N}\Delta N
\end{equation}
The equation (\ref{19}) immediately shows that the mean curvature $\tau = \frac{Tr. K_{ij}}{n}$ of $\Sigma$ is a function of only $t$, i.e. the mean curvature of each spatial slice is constant on that slice.\\

\noindent
Further, equations (\ref{20}) and (\ref{21}) imply that
\begin{equation}\label{22}
R=-n(n-1)\tau^2
\end{equation}
showing that spatial slices $\Sigma$ have non-positive scalar curvature.\\

\noindent
\textbf{Remark.} For Schwarzschild exterior space-time metric: $-(1-\frac{2m}{r})dt^2 +(1-\frac{2m}{r})^{-1}dr^2+(r^2)(d\theta^2 +sin^2 \theta d\phi^2)$, we know that the spatial slices $\Sigma$ are totally geodesic ($\tau =0$) in the space-time, $R=0$, and $N=(1-\frac{2m}{r})^{1/2}$ depends only on spatial coordinate $r$ and hence $\Delta N=0$, from equation (\ref{20}). Consequently (\ref{18}) assumes the form $R_{ij}=\frac{1}{N}(\nabla_i \nabla_j N)$. From this, it follows by a straightforward computation that the only non-zero components of the Ricci tensor of $\Sigma$ are $R^{1}_{1}=\frac{-2m}{r^3}, R^{2}_{2}=\frac{m}{r^3}, R^{3}_{3}=\frac{m}{r^3}$, which are well known.\\

\noindent
At this point, we assume that ($M,g$) is also geodesic, i.e. the acceleration vector $\textbf{A}$ vanishes. So $N$ depends only on $t$ and hence can be taken equal to 1,  by time-rescaling. Thus equation (\ref{20}) reduces to $\dot \tau = -\tau^2$ and hence integrates as $\tau = 1/t$. The use of (\ref{6}) in the foregoing equation and time integration gives $a = tX$ where $X$ is an arbitrary function of $x^k$ and can be absorbed in $\gamma_{ij}$. Consequently, the space-time metric on $M$ becomes the Lorentzian cone: $-dt^2 + t^2 \gamma_{ij}dx^i dx^j$ and $R_{ij} = -\frac{n-1}{t^2}g_{ij}$. For $dim(M)= 4$, $\Sigma$ are 3-dimensional and hence of constant negative curvature, consequently $M$ is Minkowski. Thus, it represents the expanding hyperbolic model in the Minkowski space-time (Misner, Thorne and Wheeler \cite{M-T-W}). This leads to the following result.

\begin{theorem}
If a vacuum umbilically synchronized space-time ($M,g$) is geodesic, then it is a Lorentzian cone over a negatively Einstein manifold. If, the dimension of $M$ is 4, then it represents the expanding hyperbolic model.
\end{theorem}
\noindent

\noindent
Now, for the ($n+1$)-dimensional Lorentzian cone: $-dt^2 + t^2 \gamma_{ij}dx^i dx^j$ with hyperbolic metric $\gamma_{ij}$ , we know that it is vacuum and the electric part of its Weyl tensor vanishes (\cite{Sharma-Duggal}). Motivated by this fact, we consider the spatially complete vacuum umbilically synchronized space-time for which the electric part of the Weyl tensor vanishes, i.e. $\bar{C}_{0i0j} = 0$. The geodesic case was covered in Theorem 2. For the non-geodesic case, we will show that $\Sigma$ will have geometric and topological constrains. More precisely, we prove the following result. 

\begin{theorem}
Let ($M,g$) be a spatially complete vacuum umbilically synchronized non-geodesic space-time such that the electric components of its Weyl tensor vanish. Then the spatial slices $\Sigma$ of $M$ are isometric to either (i) the product of a real line and a complete Ricci-flat Riemannian manifold, or (ii) a Euclidean space or (iii) the warped of the real line and a non-positively complete Einstein manifold. Also, $M$  cannot be spatially compact.
\end{theorem} 

\noindent
\textbf{Proof.} We have the equations (\ref{17}) and (\ref{18}) immediately decouple as
\begin{equation}\label{23}
R_{ij}=\frac{R}{n}g_{ij},
\end{equation}
and
\begin{equation}\label{24}
\nabla_i \nabla_j N=\frac{\Delta N}{n}g_{ij}.
\end{equation}
It follows from equations (\ref{23}) and (\ref{22}) that
\begin{equation*}
R_{ij}=-(n-1)\tau^2 g_{ij}
\end{equation*}
i.e. $\Sigma$ is non-positively Einstein.\\

If $\Delta N =0$ on $\Sigma$, then by a result of Tashiro \cite{Tashiro}, as $\Sigma$ is complete, it is the product of the real line and a complete Ricci-flat Riemannian manifold (because $\Sigma$ is Einstein). If $\Delta N$ is a non-zero constant on $\Sigma$, then  by a theorem of \cite{Tashiro}, $\Sigma$ is isometric to the Euclidean space $E^n$. Now we consider the case when $\Delta N$ is non-constant on $\Sigma$. As $\Sigma$ is complete, is non-positively Einstein, and $\nabla N$ is non-homothetic conformal by virtue of equation (\ref{28}), we invoke a theorem of Kanai \cite{Kanai} to conclude that $\Sigma$ is Euclidean (for $R=0$) or isometric to the warped product $R \times _{f}N$ of the real line $R$ and a complete non-positively Einstein manifold $\Sigma^*$, with warping function $f(\rho)=c_1 e^{\sqrt{-k}\rho}+c_2 e^{-\sqrt{-k}\rho}$ where $c_1$ and $c_2$ are nonnegative constants, $\rho$ is the coordinate on $R$,  and $k =\frac{R}{n(n-1)}$ for $R<0$.\\

\noindent
Now we will show that $M$ cannot be spatially compact. Assume that $M$ is spatially compact. Then $\Sigma$ is compact. Suppose $\Delta N$ is constant. Integrating it over $\Sigma$ and using divergence theorem shows that $\Delta N=0$. Hence, by Hopf's lemma, $N$ depends only on $t$, and therefore $M$ is geodesic with respect to $\textbf{n}$, contradicting the  hypothesis of the theorem. If $\Delta N$ is non-constant, then, from equation (\ref{28}) we see that $\nabla N$ is non-homothetic conformal vector field on $\Sigma$ and hence by the following result of Yano and Nagano \cite{Y-N} ``A compact Einstein manifold admitting a non-homothetic conformal vector field is isometric to the round sphere" we conclude that $\Sigma$ is isometric to a round sphere, but as noticed earlier, $\Sigma$ is non-positively Einstein, leading to a contradiction.This completes the proof.

\section{Structure Of Conformal Vector Fields} Let us recall the fact that $FLRW$-space-time admits a maximal conformal group (Maartens and Maharaj \cite{M-M}). Intrigued by this, we would like to examine a 1-parameter group of conformal motions generated by a conformal vector field $V$ defined by
\begin{equation}\label{25}
\pounds_V g=2\psi g
\end{equation}
where $\pounds$ denote the Lie-derivative operator, and $\psi$ a smooth function called the conformal scale function, on an umbilically synchronized space-time $M$. Let the conformal Killing vector field be decomposed as $V = \alpha \textbf{n}+U$ where $\alpha$ is a function on $M$ and $U$ is the component of $V$ tangential to the spatial slices $\Sigma$. The ($\textbf{n},\textbf{n}$) projection of the conformal Killing equation (\ref{25}) provides the relation
\begin{equation}\label{26}
g(\bar{\nabla}_{\textbf{n}}U,\textbf{n})=\textbf{n}\alpha-\psi.
\end{equation}
For an arbitrary vector field $X$ tangent to $\Sigma$, we take the ($X,\textbf{n}$)-projection of (\ref{25}) and use the umbilicity condition $\bar{\nabla}_X \textbf{n}=\tau X$ along with equation (\ref{26}) in order to get
\begin{equation}\label{27}
\bar{\nabla}_{\textbf{n}}U=\tau U +(\psi-\textbf{n}\alpha) \textbf{n}+N\nabla (\frac{\alpha}{N})
\end{equation}
where $\nabla$ is the spatial gradient operator.  Also, for $X,Y$ tangent to $\Sigma$, the ($X,Y$)-projection of the conformal Killing equation (\ref{25}) shows
\begin{equation}\label{28}
(\pounds_U g)(X,Y)=2(\psi-\alpha \tau)g(X,Y)
\end{equation}
and hence $U$ is conformal on spatial slices $\Sigma$. At this point, we consider the following cases.\\

\noindent
\textit{Case 1}. $V=\alpha \textbf{n}$. Hence $U=0$. Equations (\ref{26}), (\ref{27}) and (\ref{28}) boil down to (i) $\textbf{n}\alpha = \psi$, (ii) $\nabla \alpha = \alpha \nabla (ln N)$, and (iii) $\psi=\alpha \tau$. It follows from equations (\ref{6}), (i) and (iii) that $\partial_t ln \alpha=\frac{\bar{a}}{a}$ which easily integrates as $\alpha=aX$ for an arbitrary positive function $X$ of spatial coordinates. However, $X$ can be absorbed by the time-independent metric $\gamma_{ij}$ defined by (\ref{5}). Thus we have
\begin{equation}\label{29}
\alpha =a.
\end{equation}
Now (ii) gets integrated as
\begin{equation}\label{30}
\alpha = NT,
\end{equation}
where $T$ is a function of only $t$. Using this in (iii) and equation (\ref{6}) gives
\begin{equation}\label{31}
\psi = \frac{\dot a}{a}T.
\end{equation}
Consequently, $V=a\textbf{n}$ and $\psi = \frac{\dot a}{N}$. We state this finding as

\begin{proposition}
If a conformal vector field $V$ on an umbilically synchronized space-time is orthogonal to spatial slices $\Sigma$, then $V=a\textbf{n}$ and the conformal scale function $\psi = \frac{\dot a}{N}$ where $a$ is the warping function defined in the space-time metric following (\ref{5}). 
\end{proposition}

\noindent
\textbf{Remark.} This proposition gives a generalization of the natural conformal vector field $a\partial_t$ with $\psi= \dot a$ of the $FLRW$ space-time for which $N=1$ and $\textbf{n}=\partial_t$.\\

\noindent
Let us see what happens wnen $V$ is non-Killing homothetic ($\psi$ is non-zero constant) in \textit{Case 1}. The equation (\ref{31}) integrates to $a=Ye^{\psi \int(1/T)dt}$, where $Y$ is a function of only the spatial coordinates and hence can be absorbed by $a$. Thus $a=e^{\psi\int(1/T)dt}$, and hence depends only on $t$. Now, equations (\ref{29}) and (\ref{30}) show that $a=NT$. As $a$ and $T$ depend only on $t$, therefore so does $N$. By time-rescaling we can therefore take $N=1$. Consequently, $a=T$, and hence it follows from (\ref{31}) that $\dot a = \psi$ which integrates to $a=\psi t +c$ for a constant $c$. By rescaling and translating $t$ we can have $a=t$. Also, from (\ref{29}) we have $\alpha=t$. As a result, the line-element becomes $-dt^2 + t^2 \gamma_{ij}dx^i dx^j$, i.e. the Lorentzian cone, and the homothetic vector field becomes $t\partial_t$.\\

\noindent
\textit{Case 2}. $V=U$ ($\alpha = 0$) and hence tangential to $\Sigma$'s. In this case, (\ref{26}) provides $g(\bar{\nabla}_{\textbf{n}}\textbf{n},U)=\psi$. But, $\bar{\nabla}_{\textbf{n}}\textbf{n}=\textbf{A}=\nabla ln N$. Hence we get $g(\textbf{A},U)=U ln N =\psi$. From this, we obtain the following result.
\begin{theorem}Let a conformal vector field $V$ on an umbilically synchronized space-time be tangential to spatial slices $\Sigma$. Then $V$ is Killing if and only if the lapse function $N$ is constant along $V$, equivalently, $V$ is orthogonal to the acceleration vector field $\textbf{A}$.
\end{theorem}

\noindent
\textbf{Remark.} The above result is a generalization of the well known result that conformal vector fields tangential to the space-like slices of a $FLRW$ space-time are Killing \cite{M-M}, because the lapse function $N=1$ for a  $FLRW$ space-time.\\

\noindent
Finally, we find inheriting conformal vector fields on an umbilically synchronized space-time. Following Coley and Tupper \cite{Coley-Tupper}, a conformal vector field is said to be an inheriting conformal vector field, if it preserves the flow lines along the unit time-like vector field up to a function multiple.  Also, noting that $\bar{\nabla}_U \textbf{n}=\tau U$ (as $U$ is tangent to $\Sigma$) and using it in conjunction with (\ref{27}) gives $[U,\textbf{n}]=(\textbf{n}\alpha-\psi)\textbf{n}-N\nabla (\frac{\alpha}{N})$. Hence, we compute $\pounds_V \textbf{n}=[\alpha\textbf{n}+U,\textbf{n}]=-(\textbf{n}\alpha)\textbf{n}+[U,\textbf{n}]=-\psi \textbf{n}-N\nabla (\frac{\alpha}{N})$. From this, it follows that $\pounds_V \textbf{n}$ is a multiple of $\textbf{n}$ if and only if $\nabla (\frac{\alpha}{N}=0$, i.e. $\alpha= NT$ where $T$ is a function of only $t$. In this case the inheriting conformal vector field assumes the form $V=NT\textbf{n}+U= T\partial_t +U$ and $\pounds_V \textbf{n}=-\psi \textbf{n}$. Also, for $X,Y$ tangent to $\Sigma$, the ($X,Y$)-projection of the conformal Killing equation (\ref{25}) gives $(\pounds_U g)(X,Y)=2(\psi-\alpha \tau)g(X,Y)$. Summing up these findings, we  obtain the following characterization of an inhereting conformal vector field on an umbilically synchronized space-time.

\begin{theorem} A conformal vector field $V$ on an umbilically synchronized spacetime is an inheriting conformal vector field with respect to the flow lines determined by the unit timelike vector field $\textbf{n}$ if and only if $V= NT(t)\textbf{n}+U$ where $N$ is the lapse function, $T(t)$ is a function of only $t$, and $U$ is orthogonal to $\textbf{n}$.
\end{theorem}

\noindent
\textbf{Remarks.} Considering a time-like conformal vector field $V=\alpha \textbf{n}$ for a scalar function $\alpha$, we obviously see that $V$ is an inhereting conformal vector field. These vector fields arise (see Israel \cite{Israel}) as the inverse temperature function $(1/T)u$ (here $u=\textbf{n}$ and $T$ is the temperature). Their existence has been further supported by Stephani \cite{Stephani} in terms of complete exact reversible thermodynamics, and argued in favor by Tauber and Weinberg \cite{Tauber-Weinberg} in terms of the isotropy of the cosmic microwave background.

\section{Statements and Declarations}
\textbf{Data Availability}: No data was used.

\noindent
\textbf{Competing Interests}: The author states that there is no conflict of interest. \\

\end{document}